\documentclass[12pt]{iopart}

\usepackage{iopams}
\usepackage{bm}
\begin{document}

\title{The Schwartz-Soffer and more inequalities for random fields}  
\author{ C. Itoi $^1$ and Y. Sakamoto$^2$}

\address{$^1$  Department of Physics,  GS $\&$ CST,
Nihon University, Kandasurugadai, Chiyoda,  Tokyo 101-8308, Japan\\
 $^2$
 Laboratory of Physics, CST,
Nihon University, Narashinodai, Funabashi-city, Chiba 274-8501, Japan }
\vspace{10pt}

\begin{abstract} A new series of correlation inequalities for random field  spin systems is proven rigorously. 
First one corresponds to the well-known Schwartz-Soffer inequality. 
These are expected to rule out incorrect results calculated in effective theories 
and numerical studies.  
The large $N$ expansion with the replica method for random field systems as an example is checked
by these inequalities.   It is shown that several critical exponents of multiple-point  correlation functions 
at critical point satisfy obtained inequalities.    
 \end{abstract}

%
\vspace{2pc}

\section{Introduction}
The Schwartz-Soffer inequality is  a well-known useful inequality for random field spin systems \cite{SS}. 
This inequality claims that the Fourier transformed connected correlation function is bounded from the above by the square root of the corresponding disconnected one.
This relation is quite useful to rule out  incorrect results obtained by effective theories 
and numerical studies for random field spin systems.  
It is proven in a simple way with  integration by parts over the Gaussian random fields and the Cauchy-Schwarz inequality. 
The Schwartz-Soffer inequality can check  several approximation theories.
 Critical exponents calculated in the functional renormalization group 
\cite{F,MS,SMI,TT1,TT2,TT3,TT4,TBT,FK,Fe,BTT,S,TT5}, the large $N$ expansion with the replica method \cite{SMI,MY} and recent   
numerical method \cite{FM1,FMPS1,FM2,FMPS2,FMPS3}  satisfy the Schwartz-Soffer inequality. 
A few studies discuss generalization of this inequality \cite{VS}.  

In the present paper,  we provide a new series of inequalities for multiple-point  correlation functions
in spin systems with Gaussian random fields. These inequalities are obtained 
in a square interpolation method, which is a rigorous mathematical method used for Gaussian random spin models extensively \cite{G1,GT,G,C,T2,T,IU,I,I3}. 
This method  was used for the first  time to obtain Guerra's  replica symmetric bound on the free energy density of the Sherrington-Kirkpatrick model \cite{G1}.   
The usefulness of this method has attracted  attention of many mathematicians and physicists  
 since Talagrand  proved the validity of the  Parisi formula \cite{Pr} for the replica symmetry breaking  free energy density
 in the Sherrington-Kirkpatrick model \cite{SK} rigorously \cite{T2,T}.  Chatterjee generalized this method  to  evaluate bounds on
  correlation functions as well as on the free energy density  in random spin systems \cite{C}. Chatterjee's generalization is quite useful to evaluate several 
  observables \cite{IU,I,I3}. We use his method to obtain a series of inequalities.
The first one in this series is the Schwartz-Soffer inequality which gives a lower bound on a disconnected correlation function.   
The second one in the series gives an upper bound on the disconnected correlation function. 
These and other inequalities for multiple-point correlation functions  are expected to rule out incorrect results furthermore.

This paper is organized as follows. Section 2 gives a definition of  random field spin models  and our main theorem.
In Section 3, two lemmas are proven, and these enable us to prove our main theorem. 
In Section 4, several multiple-point correlation functions are calculated  in large $N$ expansion with the replica method.
It is shown  that critical exponents calculated in the large $N$ expansion with the replica method satisfy these 
inequalities. Section 5 summarizes our results.  

\section{Definitions and main theorem}
First, we define the model and functions. 
Coupling constants  in a system with quenched disorder are given by  independent and  identically distributed (i.i.d.) random variables.
We can regard a given disordered sample as a system obtained by a random sampling of these variables.
All physical quantities in such systems are functions of these random variables.
Consider a random field O($N$) invariant  Ginzburg-Landau model
on a $d$ dimensional 
 hyper cubic lattice  $\Lambda_L:= [1,L]^d \cap {\mathbb Z}^d$ whose volume is $|\Lambda_L|=L^d$.
Let  $J=(J_{x,y})_{x,y\in \Lambda_L}$ be a real symmetric matrix such that $J_{x,y}=1$, if $|x-y|=1$, otherwise $J_{x,y}=0$. 
Define  Hamiltonian as a function of $N$ dimensional spin vector configurations  $ \phi=(\phi_x^n)_{x\in\Lambda_L,  n=1,2, \cdots, N} \in ({\mathbb R}^N)^{\Lambda_L}$
and  i.i.d.standard Gaussian random variables  $g=(g_x^n)_{ n=1, 2, \cdots, N}$ by
\begin{equation}
H(\phi,g) :=-\sum_{x,y \in \Lambda_L} J_{x,y}  \phi_x \cdot \phi_y  -h \sum_{x\in \Lambda_L}g_x \cdot \phi_x,
\label{hamil}
\end{equation}
with a real constant $ h$.
Here, we define Gibbs state for the Hamiltonian.
For a positive $\beta $,  the  partition function is defined by
\begin{equation}
Z_L(\beta,h,g) :=  \int_{{\mathbb R}^{N|\Lambda_L|}} D \phi e^{ - \beta H(\phi,g)}.
\label{partition}
\end{equation}
 The measure $D\phi$ is  O($N$) invariant, for example, it is defined by
 \begin{equation}
D \phi :=C 
\prod_{x \in \Lambda_L} \prod_{n=1}^Nd \phi_x ^ne^{-u (\phi_x \cdot \phi_x -1)^2 
},
\label{phi4}
\end{equation}
 where $u > 0$ and  $C$ is a normalization constant satisfying
$$\displaystyle C^{-1} = \int_{{\mathbb R}^{N|\Lambda_L|}} D \phi.$$
The following is also possible 
\begin{equation}
D \phi =
\prod_{x \in \Lambda_L} \prod_{n=1}^Nd \phi_x ^n \delta (\phi_x \cdot \phi_x-1),
\label{Dphi} 
\end{equation}
 which can be obtained by the limit $u\to\infty$ of (\ref{phi4}).

The  expectation of a function of spin configuration $f(\phi)$ in the Gibbs state is given by
\begin{equation}
\langle f(\phi) \rangle_g =\frac{1}{Z_L(\beta,h,g)} \int_{{\mathbb R}^{N|\Lambda_L|}} D\phi f(\phi)  e^{ - \beta H(\phi,g)}.
\end{equation}

Define the following function of  $(\beta, h) \in [0,\infty) \times {\mathbb R}$ and randomness
$g=(g_x^n)_{x \in \Lambda_L, n=1,2, \cdots, N}$
\begin{equation}
\psi_L(\beta, h,g) := \frac{1}{|\Lambda_L|} \log Z_L(\beta,h,g), \\ 
\end{equation}
$-\frac{|\Lambda_L|}{\beta}\psi_L(\beta,h,g)$ is called free energy in statistical physics.
We define a function $p_L:[0,\infty) \times {\mathbb R} \rightarrow {\mathbb R}$ by
\begin{eqnarray}
p_L(\beta,h):={\mathbb E} \psi_L(\beta,h,g),
\label{FreeEner}
\end{eqnarray}
where ${\mathbb E}$ stands for the expectation over the random variables $(g_x^n)_{x \in \Lambda_L, n=1,2, \cdots, N}$.
Impose the periodic boundary condition for all variables on the lattice $\Lambda_L$ and define their Fourier transformation 
\begin{equation}
\tilde \phi_q^n := \frac{1}{\sqrt{|\Lambda_L|}} \sum_{x\in \Lambda_L} e^{-i q\cdot x} \phi_x^n, \ \  \ 
\tilde g_q^n := \frac{1}{\sqrt{|\Lambda_L|}} \sum_{x\in \Lambda_L} e^{-i q\cdot x} g_x^n,
\end{equation}
where $q \in \frac{2\pi}{L} \Lambda_L =: \Lambda_L^*$.
The random field Hamiltonian is 
\begin{equation}
 \sum_{x\in \Lambda_L}g_x \cdot \phi_x  =\sum_{q\in \Lambda_L^*} \tilde g_q \cdot \tilde \phi_{-q}.
\end{equation}
Define a connected correlation function for arbitrary  operators  $A_1, \cdots A_j$ for a positive integer $j$
 by 
 $$
 \langle A_1; \cdots ; A_j \rangle_g :=\Big[ \frac{\partial^j}{\partial b_1  \cdots \partial  b_j} \log Z_L(b, \beta, h, g) \Big]_{b_1=\cdots =b_j=0,} 
 $$
 where  the generating function is defined for $b:=(b_1, \cdots, b_j) \in {\mathbb R}^j$ by  
 $$
  Z_L(b, \beta, h, g) =  \int_{{\mathbb R}^{N|\Lambda_L|}} D \phi e^{ - \beta H(\phi,g) +\sum_{i=1} ^j A_ib_i }.
 $$
 Note that $Z_L(0, \cdots, 0, \beta, h,g)=Z_L(\beta, h, g)$,  and 
 $
 \langle A_1 ; A_2 \rangle_g := \langle A_1  A_2 \rangle_g -\langle A_1 \rangle_g \langle  A_2 \rangle_g,
 $
 for $j=2$. 
Then, we have the following main theorem.
\\ 

\noindent
{\bf Theorem } {\it Consider the random field O($N$) invariant ferromagnetic spin model. For  a  non-negative integer  $k, l$,  the following  inequalities 
for the variance of $k$-point correlation functions are valid
\begin{eqnarray*}
&&\frac{\beta^{2l } h^{2l}}{l!}\sum_{p_1 \in \Lambda_L^*} \cdots \sum_{p_{l} \in \Lambda_L ^*} 
\sum_{n_1=1} ^N \cdots \sum_{n_{l}=1}^N | {\mathbb E} W_{p_1, \cdots, p_{l}, \vec f ;g} ^{n_1, \cdots, n_{l}} |^2\\
&&\leq
{\rm  Var} \langle f_1(\phi) ; \cdots ; f_k(\phi) \rangle_g  
\leq   \beta^{2} h^{2} \hspace{-2mm} \sum_{p \in \Lambda_L ^*}
\sum_{n=1} ^N  {\mathbb E} | W_{p, \vec f,g} ^{n} |^2,
\end{eqnarray*}
 where (k+l)-point correlation function  for a sequence of complex valued  functions $\vec f := ( f_1(\phi),  \cdots , f_k(\phi))$ of a spin configuration, 
 is defined by
$$
W_{p_1, \cdots, p_{l}, \vec f ,g} ^{n_1, \cdots, n_{l}}:=\langle \tilde \phi_{p_1} ^{n_1} ; \cdots ; \tilde \phi_{p_{l}} ^{n_{l}} ; f_1(\phi); \cdots ; f_k(\phi) \rangle_g.
$$ and  the sample variance is defined by 
 $${\rm Var}   F(g) :={\mathbb E} | F(g)  |^2 -|{\mathbb E} F(g)|^2,$$
for a complex valued function $F$ of the sequence of  random fields $g$.  }\\

\noindent
{\bf Note} {\it For $k=l=1$ and $f_1(\phi) = \tilde \phi_q^m$,  Theorem gives the following upper and lower bounds on the disconnected  correlation function
 \begin{eqnarray}
({\mathbb E} \langle \tilde \phi_ q ^m; \tilde \phi_{-q}^n \rangle_g)^2 
\leq  
\beta^{-2} h^{-2}{\mathbb E} |\langle \tilde \phi_ q^m \rangle_g |^2
\label{SSieq}
\leq\sum_{p\in \Lambda_L ^*} \sum_{n=1}^N  {\mathbb E} |\langle \tilde \phi_ q ^n; \tilde \phi_{p} ^m\rangle_g |^2.
\label{SSSieq} 
\end{eqnarray} 
Note  that
$({\mathbb E} \langle \tilde \phi_ q ^m; \tilde \phi_{-q}^n \rangle_g)^2 
\leq 
\sum_{p\in \Lambda_L ^*}\sum_{n=1}^N |{\mathbb E} \langle \tilde \phi_ q ^m; \tilde \phi_{p} ^n\rangle_g |^2$
 and ${\mathbb E}\langle \tilde  \phi_{q} ^m \rangle_g =0$ for any $m$ and $q \in \Lambda_L ^*$, because of the O($N$) symmetry.
The lower bound (\ref{SSieq}) implies the Schwartz-Soffer inequality \cite{SS}.}\\


\section{
Proof
}



Theorem can be proven in terms of the square root interpolation method  used in disordered systems\cite{C,G,IU,T}. 
Let $g=( g_x^n)_{x\in\Lambda_L ,n=1,2,\cdots, N }, g' =( g_x ^n{ }')_{x \in \Lambda_L,  n=1,2,\cdots, N }$ 
be two sequences of  i.i.d. standard Gaussian variables,  and their  Fourier transformed sequences
 $\tilde g=(\tilde g_q^n)_{q\in\Lambda_L ^*,n=1,2,\cdots, N }, \tilde g' =(\tilde g_q ^n{ }')_{q\in \Lambda_L ^*,  n=1,2,\cdots, N }$
and define a function of $t \in [0,1]$ by
$$
G(t) :=  \sqrt{t} g +\sqrt{1-t} {g }',
$$
and its Fourier transform 
$$
\tilde G(t) :=  \sqrt{t} \tilde g +\sqrt{1-t} {\tilde g }'.
$$

For a sequence of functions $\vec f := ( f_1(\phi),  \cdots , f_k(\phi))$ of a spin configuration,
define a generating  function $\gamma_{\vec f} (t )$ of a parameter $s \in [0,1] $ by
\begin{equation}
\gamma_{\vec f}(t) = {\mathbb E} |{\mathbb E}' \langle f_1(\phi); \cdots ; f_k(\phi) \rangle_{G(t)} |^2,
\end{equation}
where ${\mathbb E }$ and  ${\mathbb E }'$ denote expectation
over $\tilde g$ and $\tilde g'$, respectively. This generating function 
$\gamma_{\vec f}(t)$ is an analogue to that introduced by Chatterjee \cite{C}.
This generating function is useful and utilized several studies on random spin systems\cite{IU,I,I3}. \\

\noindent
{\bf Lemma 1 }  {\it
For any $(\beta,h) \in [0,\infty) \times {\mathbb R}$, any positive integers 
$k, l$ and  a sequence  of complexvalued functions  $\vec f = ( f_1(\phi),  \cdots , f_k(\phi))$  of a spin configuration, 
 $l$-th order derivative of $\gamma_{\vec f}(t)$ is represented in the following connected correlation function for an arbitrary $t \in [0,1]$
\begin{eqnarray}
\gamma_{\vec f}^{(l)} (t)
=\beta^{2l} h^{2l} \hspace{-1.3mm} \sum_{q_1 \in \Lambda_L ^*}  \hspace{-1.3mm}  \cdots  \hspace{-1.3mm} 
\sum_{q_l \in \Lambda_L ^*}  \hspace{-1.3mm} \sum_{n_1=1} ^N  \hspace{-1.3mm} \cdots  \hspace{-1mm}  \sum_{n_l=1}^N \hspace{-1mm}
{\mathbb E} | {\mathbb E}'W_{q_1, \cdots, {q_l}, \vec f,G(t)} ^{n_1, \cdots, n_l}|^2,
\label{kth}
\end{eqnarray}
\noindent Proof. 
} 
For short hand notation,  denote 
$$
\langle \vec f \rangle_{G(t)} :=  \langle  f_1(\phi); \cdots ; f_k(\phi) \rangle_{G(t)}.
$$ 
The first derivative  of $\gamma_{\vec f}$  is calculated in integration by parts
\begin{eqnarray*}
&&\hspace{-1.5mm}\gamma_{\vec f}' (t) 
\hspace{-1.5mm}
={\mathbb E} \Big[{\mathbb E}' \langle \vec f \rangle_{G(t)}^*
{\mathbb E}' \hspace{-1mm}\sum_{p \in \Lambda_L ^*} \hspace{-1mm}\sum_{n=1} ^N\hspace{-1mm}\Big( \frac{\tilde g_p^n}{2\sqrt{t}} 
-\frac{\tilde g_p^n{}'}{2\sqrt{1-t}}  \Big) \hspace{-.8mm}\frac{\partial \langle  \vec f \rangle_{G(t)}}{ \partial {\tilde G}_{p}^n }\hspace{-.5mm} +\hspace{-.3mm}
{\rm  c.c.} \hspace{-.3mm}\Big]\\
&&\hspace{-1.5mm}= {\mathbb E}\sum_{p \in \Lambda_L ^*}\sum_{n=1} ^N\Big[  \frac{1}{2\sqrt{t}}\frac{\partial}{\partial \tilde g_p^n}{\mathbb E}'  \langle \vec f \rangle_{G(t)} ^*
{\mathbb E}' \frac{\partial \langle \vec f\rangle_{G(t)} }{ \partial {\tilde G}_{p}^n}\\&&
\hspace{1.2mm}-{\mathbb E}' \langle  \vec f \rangle_{G(t)} ^*{\mathbb E}'   \frac{1}{2\sqrt{1-t}}\frac{\partial}{\partial \tilde g_p^n{}'} 
 \frac{\partial \langle  \vec f \rangle_{G(t)} }{ \partial {\tilde G}_{p}^n } + {\rm  c.c.}  \Big]\\
&&\hspace{-1.5mm}= \sum_{p \in \Lambda_L ^*} \sum_{n=1} ^N
{\mathbb E} \Big|{\mathbb E}' \frac{\partial \langle \vec  f \rangle_{G(t)} }{ \partial {\tilde G}_{p}^n } \Big|^2\\
&&\hspace{-1.5mm}=\beta^2 h^2 \sum_{p \in \Lambda_L ^*} \sum_{n=1} ^N
 {\mathbb E} \Big|{\mathbb E}'   \langle  \tilde \phi_p^n ;f_1(\phi); \cdots ; f_k(\phi) \rangle_{G(t)}  \Big|^2 .
 \end{eqnarray*}
The formula for the $l$-th order derivative $\gamma_{\vec f}^{(l)}(t)$ 
is proven by a mathematical inductivity.
$\Box$\\

Note the positive semi-definiteness of an arbitrary order derivative $\gamma_{\vec f}^{(l)} (t)$
which implies that all order  derivative functions are monotonically increasing and convex.\\

\noindent{\bf Lemma 2} {\it  \label{2} For any $t_1 < t_2$, any non-negative integers $j, k, l$, and
a sequence  $\vec f = ( f_1(\phi),  \cdots , f_k(\phi))$,  the following inequality is valid 
$$(t_2-t_1)^l \gamma_{\vec f} ^{(j+l)}(t_1) \leq l! \gamma_{\vec f}^{(j)}(t_2).$$
Proof.}    Taylor's theorem implies that  there exists $t  \in (t_1, t_2)$, such that
\begin{eqnarray}
\gamma_{\vec f}^{(j)}(t_2) 
=\sum_{i=0}^{l-1}   \frac{1}{i!}(t_2-t_1)^j \gamma_{\vec f}^{(i+j)}(t_1) 
+  \frac{1}{l!}(t_2-t_1)^l \gamma_{\vec f} ^{(l+j)}(t). 
\end{eqnarray}
 The inequality is obvious, since each Taylor coefficient is positive semi-definite. $\Box$\\

\noindent{\bf Proof of  Theorem} Note the following  representation of the variance of the $k$-point connected correlation function
$$
 {\rm Var} \langle f_1(\phi); \cdots; f_k(\phi) \rangle_g=  \gamma_{\vec f} (1) - \gamma_{\vec f} (0).
$$  
Lemma 2 and monotonicity of $\gamma_{\vec f}'(t) $ gives
$$
\frac{ \gamma_{\vec f}^{(l)}(0)}{l!} \leq  \gamma_{\vec f} ^{(l)}(0) \int_0 ^1 \frac{t^{l-1}}{(l-1)!} dt
  \leq  \int_0^1  \gamma_{\vec f} '(t) dt \leq  \gamma_{\vec f}' (1),
$$
 for a positive integer $l$.
This enables us to prove  Theorem.  $\Box$

\section{Large $N$ expansion with the replica method}

Here, we check whether a critical exponent calculated in the large $N$ expansion with the replica method obtained in Ref. \cite{SMI}
satisfies the correlation inequality (\ref{SSSieq}).
Consider the model defined by the partition  function (\ref{partition}) with O($N$) invariant  Hamiltonian (\ref{hamil})  and  measure (\ref{Dphi}) for 
$d > 4$ to study critical phenomenon in random field O($N$) spin model. 
Here, for the purpose of the field-theoretical description, 
we redefine the lattice as $\Lambda_L:=(-L/2,L/2]^d\cap{\mathbb Z}^d$. 
The boundary condition for all variables on the lattice $\Lambda_L$ 
remains periodic, and their Fourier transformation are redefined as 
\begin{eqnarray}
{\tilde{\phi}}_q^n:=\sum_{x\in\Lambda_L}e^{-iq\cdot x}\phi_x^n,\,\,\,\,\,\,
{\tilde{g}}_q^n:=\sum_{x\in\Lambda_L}e^{-iq\cdot x}g_x^n,
\end{eqnarray}
where $q\in\frac{2\pi}{L}\Lambda_L=:\Lambda_L^*$. 
In order to compute the expectation over the Gaussian random variables 
$(g_x^n)_{x\in\Lambda_L,n=1,2,\cdots,N}$ in (\ref{FreeEner}) 
by use of the replica method,  
we introduce $r$ copies of the original spins: 
$(\phi_{x,a}^n)_{x\in\Lambda_L,n=1,2,\cdots,N,a=1,2,\cdots,r}$. 
The expectation in (\ref{FreeEner}) is calculated via the following relation: 
\begin{eqnarray}
{\mathbb E}\psi(\beta,h,g)
=\frac{1}{|\Lambda_L|}{\mathbb E}\log Z_L(\beta,h,g)
=\frac{1}{|\Lambda_L|}\lim_{r\searrow 0}
\frac{{\mathbb E}Z_L(\beta,h,g)^{r}-1}{r},
\end{eqnarray}
where ${\mathbb E}Z_L(\beta,h,g)^{r}$ and the limit $r\searrow0$ are called 
the replica partition function  and the replica limit, respectively.  The expectation value of the function  $f(\phi)$ of 
spin configurations can be calculated in the following replica limit, 
\begin{eqnarray}
&&{\mathbb E} \langle  f(\phi) \rangle_g =  {\mathbb E}  \frac{1}{Z_L(\beta, h, g)}  \int_{{\mathbb R}^{N|\Lambda_L|}} D\phi f(\phi) e^{-\beta H(\phi,g)} \nonumber  \\
&=&\lim_{r\searrow 0} {\mathbb E}  \frac{1}{Z_L(\beta, h, g)^r}  \int_{{\mathbb R}^{rN|\Lambda_L|}} \prod_{a=1} ^r D\phi_a f(\phi_b) e^{-\beta\sum_{a=1}^r  H(\phi_a ,g)}.
\end{eqnarray}
Two different replicas of spin configurations $\phi_a, \phi_b$ enable us to calculate the sample expectation of the product 
of two Gibbs expectations of two functions $f_1(\phi), f_2(\phi)$ of spin configuration.
\begin{eqnarray}
\fl &&{\mathbb E} \langle  f_1(\phi) \rangle_g \langle  f_2(\phi) \rangle_g \nonumber  \\
\fl &=&  {\mathbb E}  \frac{1}{Z_L(\beta, h, g)}  \int_{{\mathbb R}^{N|\Lambda_L|}} D\phi f_1(\phi) e^{-\beta H(\phi,g)} 
 \frac{1}{Z_L(\beta, h, g)}  \int_{{\mathbb R}^{N|\Lambda_L|}} D\phi  f_2(\phi) e^{-\beta H(\phi,g)}\nonumber  \\
\fl &=&\lim_{r\searrow 0} {\mathbb E}  \frac{1}{Z_L(\beta, h, g)^r}  \int_{{\mathbb R}^{rN|\Lambda_L|}} \prod_{a=1} ^r D\phi_a f_1(\phi_1) 
f_2(\phi_2) e^{-\beta\sum_{a=1}^r  H(\phi_a ,g)}.
\end{eqnarray}
This replica method is believed to be useful, and has been employed to calculate observables  
extensively in statistical physics, although the replica limit is not mathematically  rigorous. 
There have been many results  calculated in effective theories using the replica method.  
At least, we have to check them in rigorous inequalities.  
After integrating over the Gaussian random variables 
$(g_x^n)_{x\in\Lambda_L,n=1,2,\cdots,N}$, 
we obtain the following expression for the replica partition function: 
\begin{eqnarray}
&&{\mathbb E}Z_L(\beta,h,g)^{r}
=e^{r|\Lambda_L|\beta Jd}\int\left(
\prod_{x\in\Lambda_L}\prod_{a=1}^{r}
\sqrt{N}d\phi_{x,a}\delta({\phi_{x,a}}^2-1)
\right)
e^{-\beta H_{\rm rep}},
\\
&&\beta H_{\rm rep}
=\frac{\beta}{2}
\sum_{x\in\Lambda_L}\sum_{a,b=1}^{r}
\phi_{x,a}
(-J{\hat \Delta}_x\delta_{a,b}-\beta\Delta_{\rm G})
\phi_{x,b}.
\end{eqnarray}
Here, we have redefined $J_{x,y}$ as $J_{x,y}=J/2$ $(J>0)$, 
if $|x-y|=1$, otherwise $J_{x,y}=0$. 
${\hat \Delta}_x$ denotes the lattice Laplacian in the Euclidean space, 
which is represented by 
${\hat \Delta}_q=2\sum_{\mu=1}^d(\cos q_{\mu}-1)$
in the momentum representation. 
$\Delta_{\rm G}$ denotes the strength of the Gaussian random variables 
$(g_x^n)_{x\in\Lambda_L,n=1,2,\cdots,N}$ such that 
$h^2{\mathbb E}g_x^ng_y^m=\Delta_{\rm G}\delta_{x,y}\delta_{n,m}$. 
Integrating over the replicated spin variables 
$(\phi_{x,a}^n)_{x\in\Lambda_L,n=1,2,\cdots,N,a=1,2,\cdots,r}$ 
after introducing the auxiliary variable 
$\lambda_{a x}\in{\mathbb R}$ 
to rewrite $\delta({\phi_{x,a}}^2-1)$ as 
\begin{eqnarray}
\delta({\phi_{x,a}}^2-1)
=\int_{-\infty}^{\infty}\frac{\beta d\lambda_{a x}}{4\pi}
e^{-\beta i \lambda_{a x}({\phi_{x,a}}^2-1)/2},
\end{eqnarray}
the replica partition function becomes 
\begin{eqnarray}
\fl {\mathbb E}Z_L(\beta,h,g)^{r}
&=&e^{Nr|\Lambda_L|\beta Jd}
\left(\frac{N\beta}{4\pi}\right)^{r|\Lambda_L|}
\left(\frac{2\pi}{\beta}\right)^{Nr|\Lambda_L|/2}
\int\left(
\prod_{x\in\Lambda_L}\prod_{a=1}^{r}
d\lambda_{a x}
\right)
e^{-S_{\rm eff}},
\end{eqnarray}
\begin{eqnarray}
S_{\rm eff}
=\frac{N}{2}
\sum_{x\in\Lambda_L}
\langle x|
\Tr\ln(-J{\hat \Delta}_x{\bm 1}_{r}+{\bm \chi})
|x \rangle
-\frac{N\beta}{2}\sum_{x\in\Lambda_L}\sum_{a=1}^{r}
i\lambda_{a x},
\end{eqnarray}
where ${\bm 1}_{r}$ is an $r \times r$ unit matrix, 
${\bm \chi}$ is an $r \times r$ symmetric matrix with 
\begin{eqnarray}
\chi_{ab x}
=i\lambda_{a x}\delta_{a,b}-\beta\Delta_{\rm G}. 
\end{eqnarray}
Here we have redefined the parameters as follows: 
\begin{eqnarray}
\frac{\beta}{N}\rightarrow\beta,
\nonumber\\
N\Delta_{\rm G}\rightarrow\Delta_{\rm G},
\nonumber
\end{eqnarray}
to keep $\beta/N$ and $N\Delta_{\rm G}$ with finite 
in the large $N$ limit. 
We should note that 
the relation between $\beta^2 h^2$ 
which is appeared in the previous sections 
and $\beta\Delta_{\rm G}$ 
which is introduced in this section is given by 
$\beta^2 h^2\leftrightarrow\beta\Delta_{\rm G}$.

\subsection{Replica-symmetric saddle-point equation and expansion of $S_{\rm eff}$ around replica-symmetric saddle point}

Note that the replica symmetry breaking (RSB) does not occur in the random field Ising model 
for almost all $(\beta, h) \in (0, \infty) \times {\mathbb R}$ \cite{C2}.  It is believed that RSB does not occur  either in the random field O($N$) spin model 
\cite{F,SMI,TT1,TT2,TT3,TT4,TBT,FK,Fe,S,TT5}. 
Here, we assume replica symmetry to calculate correlation functions.  
The saddle-point equation is obtained by diffentiating $S_{\rm eff}$ by $i\lambda_{a x}$ as follows: 
\begin{eqnarray}
\frac{\delta S_{\rm eff}}{\delta i\lambda_{a x}}
=\frac{N}{2}\Biggl\langle x\Biggr|
\left(
\frac{1}{-J{\hat{\Delta}}_x{\bm 1}_{r}+{\bm \chi}}
\right)_{aa}
\Biggl|x\Biggr\rangle
-\frac{N\beta}{2}=0.
\end{eqnarray}
We assume the replica symmetry
\begin{eqnarray}
i\lambda_{a x}=m^2,
\end{eqnarray}
and write the replica-symmetric solution formally as 
\begin{eqnarray}
{\bar{\chi}}_{ab}=m^2\delta_{a,b}-\beta \Delta_{\rm G}. 
\end{eqnarray}
In this assumption, the propagator is written in the momentum representation as 
\begin{eqnarray}
\Biggl\langle k\Biggr|
\left(
\frac{1}{-J{\hat{\Delta}}_x{\bm 1}_{r}+{\bar{\bm \chi}}}
\right)_{ab}
\Biggl|k \Biggr\rangle
&=&\frac{1}{-J{\hat{\Delta}}_k+m^2}\delta_{a,b}+\frac{\beta\Delta_{\rm G}}{(-J{\hat{\Delta}}_k+m^2)^2}
\nonumber\\
&=:&G_{0k}^{\rm c}\delta_{a,b}+(\beta\Delta_{\rm G})G_{0k}^{\rm d}
=:G_{0k}^{ab}.
\end{eqnarray}
Then, the saddle-point equation becomes 
\begin{eqnarray}
\beta=\frac{1}{|\Lambda_L|}\sum_{k\in\Lambda_L^*}\frac{1}{-J{\hat{\Delta}}_k+m^2}
+(\beta\Delta_{\rm G})\frac{1}{|\Lambda_L|}\sum_{k\in\Lambda_L^*}\frac{1}{(-J{\hat{\Delta}}_k+m^2)^2}.
\end{eqnarray}

We put
\begin{eqnarray}
\chi_{ab x}={\bar{\chi}}_{ab}+i\epsilon_{a,x}\delta_{a,b}
=:{\bar{\chi}}_{ab}+\delta\chi_{ab x},
\end{eqnarray}
and expand $S_{\rm eff}$ up to the second order of $\delta\chi_{ab x}$. 
The second-order term of $\delta\chi_{ab x}$ for $S_{\rm eff}$ becomes 
\begin{eqnarray}
\delta^2S_{\rm eff}
&=&-\frac{N}{4}\sum_{x\in\Lambda_L}\Biggl\langle x\Biggr|
\Tr \frac{1}{-J{\hat{\Delta}}_x{\bm 1}_{r}+{\bar{\bm \chi}}}\delta{\bm \chi}
\frac{1}{-J{\hat{\Delta}}_x{\bm 1}_{r}+{\bar{\bm \chi}}}\delta{\bm \chi}
\Biggl|x\Biggr\rangle
\nonumber\\
&=&\frac{N}{4}\frac{1}{|\Lambda_L|}\sum_{k\in\Lambda_L^*}\sum_{a,b=1}^{r}
{\tilde \epsilon}_{a,k}{\tilde \epsilon}_{b,-k}\Pi_{ab k}. 
\end{eqnarray}
Here, $\Pi_{ab k}$ is 
\begin{eqnarray}
\Pi_{ab k}
&:=&\frac{1}{|\Lambda_L|}\sum_{q\in\Lambda_L^*}G_{0k-q}^{ab}G_{0q}^{ba}
\label{polarization}
\nonumber\\
&=&
[(A*A)_k+(A*B)_k+(B*A)_k]\delta_{a,b}+(B*B)_k,
\end{eqnarray}
\begin{eqnarray}
(A*A)_k
&:=&\frac{1}{|\Lambda_L|}\sum_{q\in\Lambda_L^*}G_{0k-q}^{\rm c}G_{0q}^{\rm c},
\\
(A*B)_k
&:=&(\beta\Delta_{\rm G})\frac{1}{|\Lambda_L|}\sum_{q\in\Lambda_L^*}G_{0k-q}^{\rm c}G_{0q}^{\rm d},
\\
(B*A)_k
&:=&(\beta\Delta_{\rm G})\frac{1}{|\Lambda_L|}\sum_{q\in\Lambda_L^*}G_{0k-q}^{\rm d}G_{0q}^{\rm c},
\\
(B*B)_k
&:=&(\beta\Delta_{\rm G})^2\frac{1}{|\Lambda_L|}\sum_{q\in\Lambda_L^*}G_{0k-q}^{\rm d}G_{0q}^{\rm d}. 
\end{eqnarray}

\subsection{Calculation of multi-point correlation functions}
Here, we calculate the multi-point correlation functions up to the second order of the perturbation. 
For simplicity, we put $J=1$. 
In the thermodynamic limit $L\nearrow\infty$, 
the lattice Laplacian at criticality 
and the summation over $q\in\Lambda_L^*$ become 
\begin{eqnarray}
&&{\hat{\Delta}}_x=\sum_{\mu=1}^d\frac{\partial^2}{\partial x_{\mu}^2}=:\partial^2,
\quad{\hat{\Delta}}_k=-k^2,
\\
&&\frac{1}{|\Lambda_L|}\sum_{q\in\Lambda_L^*}
\rightarrow
\prod_{\mu=1}^d\left(\int_{-\pi}^{\pi}\frac{dq_{\mu}}{2\pi}\right)
=:\int_{[-\pi,\pi]^d}\frac{d^dq}{(2\pi)^d}=:\int_q.
\end{eqnarray}
Then, (\ref{polarization}) is 
\begin{eqnarray}
\Pi_{ab k}
=(c_0+c_1k^{d-4}+c_2k^{d-6})\delta_{a,b}+c_3k^{d-8},
\end{eqnarray}
for $m^2=0$, where 
\begin{eqnarray}
c_0
&=&
\int_q\frac{1}{q^4},
\\
c_1
&\simeq&
\frac{1}{(4\pi)^{d/2}}
\frac{\Gamma\left(\frac{6-d}{2}\right)\left[2\Gamma\left(\frac{d-2}{2}\right)^2-\frac{1}{2}\Gamma\left(\frac{d-4}{2}\right)^2\right]}{\Gamma(d-4)},
\\
c_2
&\simeq&
\frac{2(\beta\Delta_{\rm G})}{(4\pi)^{d/2}}
\frac{\Gamma\left(\frac{6-d}{2}\right)\Gamma\left(\frac{d-2}{2}\right)\Gamma\left(\frac{d-4}{2}\right)}{\Gamma(d-3)},
\\
c_3
&\simeq&
\frac{(\beta\Delta_{\rm G})^2}{(4\pi)^{d/2}}
\frac{\Gamma\left(\frac{8-d}{2}\right)\Gamma\left(\frac{d-4}{2}\right)^2}{\Gamma(d-4)}.
\end{eqnarray}

\subsubsection{Two-point correlation function}

In order to evaluate the two-point correlation functions 
${\mathbb E}\langle{\tilde\phi}_q^m;{\tilde\phi}_{-q}^n\rangle_g$ 
and ${\mathbb E}\langle{\tilde\phi}_q^m\rangle_g\langle{\tilde\phi}_{-q}^n\rangle_g$, 
we compute the following correlation function up to the second order of ${\tilde \epsilon}_{a,k}$: 
\begin{eqnarray}
G_q^{a_1a_2}\delta_{m,n}
&:=&\frac{\delta_{m,n}}{{\cal Z}_{\epsilon}}
\left(\int\prod_{k\in\Lambda_L^*}\prod_{a=1}^{r}d{\tilde \epsilon}_{a,k}\right)
\Biggl\langle q\Biggr|\left(
\frac{1}{-\partial^2{\bm 1}_{r}+{\bm \chi}}
\right)_{a_1a_2}\Biggl|q\Biggr\rangle e^{-\delta^2S_{\rm eff}}
\nonumber\\
&\simeq&\left[G_{0q}^{a_1a_2}
-\sum_{b_1,b_2=1}^{r}G_{0q}^{a_1b_1}
\int_{k}G_{0q-k}^{b_1b_2}\langle{\tilde \epsilon}_{b_1,k}{\tilde \epsilon}_{b_2,-k}\rangle_{\epsilon}
G_{0q}^{b_2a_2}\right]\delta_{m,n},
\end{eqnarray}
where ${\cal Z}_{\epsilon}$ 
and $\langle{\tilde \epsilon}_{b_1,k}{\tilde \epsilon}_{b_2,-k}\rangle_{\epsilon}$ are defined by 
\begin{eqnarray}
{\cal Z}_{\epsilon}
:=\left(\int\prod_{k\in\Lambda_L^*}\prod_{a=1}^{r}d{\tilde \epsilon}_{a,k}\right)
e^{-\delta^2S_{\rm eff}},
\end{eqnarray}
\begin{eqnarray}
\langle{\tilde \epsilon}_{b_1,k}{\tilde \epsilon}_{b_2,-k}\rangle_{\epsilon}
:=\frac{1}{{\cal Z}_{\epsilon}}
\left(\int\prod_{k\in\Lambda_L^*}\prod_{a=1}^{r}d{\tilde \epsilon}_{a,k}\right)
{\tilde \epsilon}_{b_1,k}{\tilde \epsilon}_{b_2,-k}
e^{-\delta^2S_{\rm eff}}.
\end{eqnarray}
In $4<d<6$ and in low momentum, 
$\langle{\tilde \epsilon}_{b_1,k}{\tilde \epsilon}_{b_2,-k}\rangle_{\epsilon}$ becomes 
\begin{eqnarray}
\langle{\tilde \epsilon}_{b_1,k}{\tilde \epsilon}_{b_2,-k}\rangle_{\epsilon}
\simeq\frac{2}{Nc_2}\left(\frac{1}{{k}^{d-6}}\delta_{b_1,b_2}-(\beta\Delta_{\rm G})\frac{6-d}{2}\frac{1}{{k}^{d-4}}\right)
=:(\Pi^{-1})_{k}^{b_1b_2}.
\end{eqnarray}
Thus, we get the following expression for $G_q^{\alpha\beta}$: 
\begin{eqnarray}
G_q^{a_1a_2}
\simeq\frac{1}{q^2}\left(1+\frac{d-4}{N}\log q\right)\delta_{a_1,a_2}
+\frac{\beta\Delta_{\rm G}}{q^4}\left(1+\frac{d-4}{N}\log q\right).
\label{2pt}
\end{eqnarray}

\subsubsection{Disconnected four-point correlation function}

In order to evaluate the correlation function 
$\sum_{p\in\Lambda_L^*}{\mathbb E}|\langle{\tilde\phi}_q^m;{\tilde\phi}_{p}^n\rangle_g|^2$, 
we compute the following correlation function up to the second order of ${\tilde \epsilon}_{\alpha,k}$: 
\begin{eqnarray}
G_q^{a_1a_2a_3a_4}\delta_{m,n}
&:=&\frac{\delta_{m,n}}{{\cal Z}_{\epsilon}}
\left(\int\prod_{k\in\Lambda_L^*}\prod_{a=1}^{r}d{\tilde \epsilon}_{a,k}\right)
\int_p
\Biggl\langle q\Biggr|\left(
\frac{1}{-\partial^2{\bm 1}_{r}+{\bm \chi}}
\right)_{a_1a_2}\Biggl|p\Biggr\rangle
\nonumber\\
&&\times
\Biggl\langle p\Biggr|\left(
\frac{1}{-\partial^2{\bm 1}_{r}+{\bm \chi}}
\right)_{a_3a_4}\Biggl|q\Biggr\rangle
e^{-\delta^2S_{\rm eff}}
\nonumber\\
&\simeq&\left[
G_{0q}^{a_1a_2}G_{0q}^{a_3a_4}
-\sum_{b_1,b_2=1}^{r}G_{0q}^{a_1a_2}G_{0q}^{a_3b_1}
\int_{k}G_{0q-k}^{b_1b_2}(\Pi^{-1})_k^{b_1b_2}G_{0q}^{b_2a_4}
\right.
\nonumber\\
&&\left.
-\sum_{b_1,b_2=1}^{r}G_{0q}^{a_1b_1}
\int_{k}G_{0q-k}^{b_1b_2}(\Pi^{-1})_k^{b_1b_2}
G_{0q}^{b_2a_2}G_{0q}^{a_3a_4}
\right.
\nonumber\\
&&\left.
-\sum_{b_1,b_2=1}^{r}G_{0q}^{a_1b_1}
\int_{k}G_{0q-k}^{b_1a_2}(\Pi^{-1})_k^{b_1b_2}
G_{0q}^{a_3b_2}G_{0q}^{b_2a_4}
\right]\delta_{m,n}.
\end{eqnarray}
The contribution of $G_q^{a_1a_2a_3a_4}$ to the correlation function 
$\sum_{p\in\Lambda_L^*}{\mathbb E}|\langle{\tilde\phi}_q^m;{\tilde\phi}_{p}^n\rangle_g|^2$
originates from the terms which are proportional to $\delta_{a_1,a_2}\delta_{a_3,a_4}$. 
Thus, we get the following expression for $G_q^{a_1a_2a_3a_4}$: 
\begin{eqnarray}
G_q^{a_1a_2a_3a_4}
\simeq\frac{1}{q^4}\left(1+\frac{d-4}{N}\log q\right)\delta_{a_1,a_2}\delta_{a_3,a_4}.
\label{4pt}
\end{eqnarray}

\subsubsection{Connected four-point correlation function}
In order to evaluate the correlation function 
$$\sum_{p_1,p_2,p_3\in\Lambda_L^*}\sum_{n_1,n_2,n_3=1}^{N}
|{\mathbb E}\langle{\tilde\phi}_{p_1}^{n_1};{\tilde\phi}_{p_2}^{n_2};{\tilde\phi}_{p_3}^{n_3};{\tilde\phi}_q^m\rangle_g|^2,$$ 
we compute the following correlation function up to the second order of ${\tilde \epsilon}_{a,k}$: 
\begin{eqnarray}
&&\sum_{p_1,p_2,p_3\in\Lambda_L^*}\sum_{n_1,n_2,n_3=1}^{N}
G_{p_1,p_2,p_3,q}^{a_1a_2a_3a_4}
G_{q,p_3,p_2,p_1}^{a_4a_3a_2a_1}\delta_{n_1,n_2}\delta_{n_3,m}
\nonumber\\
&&=N\sum_{p_1,p_2,p_3\in\Lambda_L^*}
G_{p_1,p_2,p_3,q}^{a_1a_2a_3a_4}
G_{q,p_3,p_2,p_1}^{a_4a_3a_2a_1}.
\end{eqnarray}
\begin{eqnarray}
G_{p_1,p_2,p_3,q}^{a_1a_2a_3a_4}
&:=&\frac{1}{{\cal Z}_{\epsilon}}
\left(\int\prod_{k\in\Lambda_L^*}\prod_{a=1}^{r}d{\tilde \epsilon}_{a,k}\right)
\Biggl\langle p_1\Biggr|\left(
\frac{1}{-\partial^2{\bm 1}_{r}+{\bm \chi}}
\right)_{a_1a_2}\Biggl|p_2\Biggr\rangle
\nonumber\\
&&\times
\Biggl\langle p_3\Biggr|\left(
\frac{1}{-\partial^2{\bm 1}_{r}+{\bm \chi}}
\right)_{a_3a_4}\Biggl|q\Biggr\rangle
e^{-\delta^2S_{\rm eff}}
\nonumber\\
&\simeq&
G_{0p_1}^{a_1a_2}G_{0p_3}^{a_3a_4}\delta_{p_1,p_2}\delta_{p_3,q}
\nonumber\\
&&
-\sum_{b_1,b_2=1}^{r}G_{0p_1}^{a_1a_2}G_{0p_3}^{a_3b_1}
\int_{k}G_{0k}^{b_1b_2}(\Pi^{-1})_{p_3-k}^{b_1b_2}G_{0q}^{b_2a_4}
\delta_{p_1,p_2}\delta_{p_3,q}
\nonumber\\
&&
-\sum_{b_1,b_2=1}^{r}G_{0p_1}^{a_1b_1}
\int_{k}G_{0k}^{b_1b_2}(\Pi^{-1})_{p_1-k}^{b_1b_2}
G_{0p_2}^{b_2a_2}G_{0p_3}^{a_3a_4}
\delta_{p_1,p_2}\delta_{p_3,q}
\nonumber\\
&&
-\sum_{b_1,b_2=1}^{r}G_{0p_1}^{a_1b_1}
G_{0p_2}^{b_1a_2}(\Pi^{-1})_{p_1-p_2}^{b_1b_2}
G_{0p_3}^{a_3b_2}G_{0q}^{b_2a_4}\delta_{p_1-p_2,q-p_3}.
\end{eqnarray}
The leading contribution of 
$G_{p_1,p_2,p_3,q}^{a_1a_2a_3a_4}G_{q,p_3,p_2,p_1}^{a_4a_3a_2a_1}$
to the correlation function 
$$\sum_{p_1,p_2,p_3\in\Lambda_L^*}\sum_{n_1,n_2,n_3=1}^{N}
|{\mathbb E}\langle{\tilde\phi}_{p_1}^{n_1};{\tilde\phi}_{p_2}^{n_2};{\tilde\phi}_{p_3}^{n_3};{\tilde\phi}_q^m\rangle_g|^2$$
originates from the terms which are proportional to 
$\delta_{a_1,a_2}\delta_{a_1,a_3}\delta_{a_3,a_4}$, $\delta_{p_1-p_2,q-p_3}$ 
and $(\beta\Delta_{\rm G})^{-2}$. 
Calculating 
$$\sum_{p_1,p_2,p_3\in\Lambda_L^*}
G_{p_1,p_2,p_3,q}^{a_1a_2a_3a_4}G_{q,p_3,p_2,p_1}^{a_4a_3a_2a_1},$$ 
we have 
\begin{eqnarray}
\sum_{p_1,p_2,p_3\in\Lambda_L^*}
G_{p_1,p_2,p_3,q}^{a_1a_2a_3a_4}G_{q,p_3,p_2,p_1}^{a_4a_3a_2a_1}
\simeq
\left(
\frac{4-d}{4(\beta\Delta_{\rm G})^2N^2}
\frac{\log q}{q^4}
\right)
\delta_{a_1,a_2}\delta_{a_1,a_3}\delta_{a_3,a_4}.
\label{8pt}
\end{eqnarray}

\subsection{Critical exponents and check of theorems and inequalities}
Assume the following  asymptotic form of  correlation functions for small wave number $q$  
\begin{equation}
{\mathbb E} \langle \tilde \phi_ q ^m ; \tilde \phi_{-q}^n \rangle_g \simeq  \frac{\delta_{m,n} }{q^{2-\eta}}, \ \ \
{\mathbb E} \langle \tilde \phi_ q ^m \rangle_g \langle \tilde \phi_{-q}^n \rangle_g \simeq  \frac{\delta_{m,n}}{q^{4-\bar \eta}}.
\end{equation} 
The Schwartz-Soffer inequality (\ref{SSieq}) imposes 
\begin{equation}
2 \eta \geq \bar \eta.
\end{equation} 
These critical exponents $\eta$ and $\bar \eta$ calculated in several functional 
renormalization group calculations \cite{F,MS,SMI,TT1,TT2,TT3,TT4,TBT,FK,Fe,BTT,S,TT5}
, 
 large $N$ expansion studies with the replica method \cite{SMI,MY}
 and   recent numerical studies 
\cite{
FM1,FMPS1,FM2,FMPS2,FMPS3} 
satisfy this inequality.  
According to (\ref{2pt}) 
in the leading order of the large $N$ expansion with the replica method \cite{SMI}, these are
  \begin{eqnarray}
&&{\mathbb E} \langle \tilde \phi_ q ^m ; \tilde \phi_{-q}^n \rangle_g \simeq  \frac{\delta_{m,n} }{q^{2}} \Big(1 + \frac{d-4}{N} \log q \Big),  \\
&&{\mathbb E} \langle \tilde \phi_ q ^m \rangle_g \langle \tilde \phi_{-q}^n \rangle_g \simeq  \frac{\delta_{m,n}}{q^{4}} \Big(1+\frac{d-4}{N} \log q \Big).
\end{eqnarray} 
Then, this expansion gives  the correlation exponents $\eta$ and $\bar \eta$ 
 \begin{equation}
 \eta = \frac{d-4}{N}, \ \ \   \bar \eta = \frac{d-4}{N},
 \label{etabareta}
 \end{equation}
for the $d$-dimensional random field O($N$) spin model \cite{SMI}. These values in (\ref{etabareta}) are consistent with results obtained in
the functional renormalization group \cite{SMI,TT1,TT2,TT3,TT4,TBT,Fe,TT5}. 
Note that  $\eta=\bar \eta$ satisfies the Schwartz-Soffer inequality (\ref{SSieq}).
This identity is well-known as the dimensional reduction which claims that the critical exponents of a random field spin 
system in dimension 
$d$ are identical to those of the corresponding spin system without random field in dimension $d -2$  \cite{AIM,PS}.
Parisi and Sourlas conjectured the dimensional reduction in  the argument of the hidden supersymmetry  \cite{PS}. 
Although the dimensional reduction and the supersymmetry conjectures 
 fail in dimensions less than four, 
its validity near six dimensions or for large $N$ is discussed still \cite{F, MS,SMI,TT1,TT2,TT3,TT4,TBT,Fe,TT5,MY,
FM1,FMPS1,FM2,FMPS2,FMPS3,FMPPS}.   
In addition to this result,  consider another critical exponent $\eta'$ of  the following correlation function 
\begin{equation}
\sum_{p \in \Lambda_L^* } {\mathbb E} |\langle \tilde \phi_ q ^m ; \tilde \phi_{p}^n \rangle_g |^2 \simeq \frac{\delta_{m,n}}{q^{4-\eta'} }.  
\end{equation}
According to the (\ref{4pt}) 
in the large $N$ expansion with the replica method,
$$\eta' = \frac{d-4}{N}$$
 is obtained.  
This result satisfies another inequality (\ref{SSSieq}).

The connected 4-point correlation function satisfies the inequality
for $k=1, l=3$ and $f_1(\phi) = \tilde \phi_q^m$ given by  Theorem.  
\begin{eqnarray}
\fl &&\sum_{p_1 \in \Lambda_L ^*}  \sum_{p_{2} \in \Lambda_L^*} \sum_{p_{3} \in \Lambda_L^*} 
\sum_{n_1=1} ^N\sum_{n_2=1} ^N \sum_{n_3=1}^N | {\mathbb E} 
\langle \tilde \phi_{p_1} ^{n_1} ; \tilde \phi_{p_2} ^{n_2}; \tilde \phi_{p_{3}} ^{n_{3}} ; \tilde \phi_q^m \rangle_g |^2 
\leq 3! \beta^{-6} h^{-6} {\mathbb E}| 
\langle \tilde \phi_q^m \rangle_g |^2.
\label{1-3}
\end{eqnarray}
According to (\ref{8pt}), the left hand side can be calculated in the large $N$ expansion
\begin{eqnarray}
\fl &&\sum_{p_1 \in \Lambda_L ^*}  \sum_{p_{2} \in \Lambda_L^*} \sum_{p_{3} \in \Lambda_L^*} 
\sum_{n_1=1} ^N\sum_{n_2=1} ^N \sum_{n_3=1}^N | {\mathbb E} 
\langle \tilde \phi_{p_1} ^{n_1} ; \tilde \phi_{p_2} ^{n_2}; \tilde \phi_{p_{3}} ^{n_{3}} ; \tilde \phi_q^m \rangle_g |^2 
\simeq 
\frac{4-d}{4 \beta^4 h^4 N} \frac{\log q}{q^4}.
\end{eqnarray}
Since the right hand side in the large $N$ expansion is
$$
3! \beta^{-6} h^{-6} {\mathbb E}|  \langle \tilde \phi_q^m \rangle_g |^2 \simeq \frac{3! }{\beta^4 h^4} \frac{1}{q^4} \Big(1+ \frac{d-4}{N} \log q \Big),
$$
these satisfy the inequality (\ref{1-3}).


The wave number dependent  susceptibility  can be represented in terms of correlation function
\begin{equation}
\tilde \chi^{m,n} (q,g) := \langle \tilde \phi_q^m; \tilde \phi_{-q}^n \rangle_g.
\end{equation}
In the large $N$ expansion, the variance of the susceptibility and a correlation function are obtained
\begin{eqnarray}
\fl &&{\rm Var} \tilde \chi ^{m,n}(q,g) =  {\mathbb E} |\langle \tilde \phi_q^m; \tilde \phi_{-q}^n\rangle_g |^2 -  |{\mathbb E} \langle \tilde \phi_q^m; \tilde \phi_{-q}^n \rangle_g|^2 
\simeq  \frac{\delta_{m,n}}{q^{4} }(\eta'-2\eta) \log q, \\
\fl &&\frac{\beta^4h^4}{2}\sum_{p_1 \in \Lambda_L ^*} 
\sum_{p_{2} \in \Lambda_L^*} 
\sum_{n_1=1} ^N\sum_{n_2=1} ^N 
 | {\mathbb E} 
\langle \tilde \phi_{p_1} ^{n_1} ; \tilde \phi_{p_2} ^{n_2}; \tilde \phi_q ^m ; \tilde \phi_{-q}^n \rangle_g |^2 
\simeq\frac{4-d}{8N} \frac{\delta_{m,n}}{q^{4} } \log q,
\end{eqnarray}
The following inequalities are obtained by $j=0, k=2, l=1$ $f_1 = \tilde \phi_q^m, f_2 = \tilde \phi_{-q}^n$  in Theorem and 
Lemma 2. These give variance inequalities for the susceptibility 
\begin{eqnarray}
\fl &&\frac{\beta^4h^4}{2}\hspace{-2mm}\sum_{p_1 \in \Lambda_L ^*} 
\sum_{p_{2} \in \Lambda_L^*} 
\sum_{n_1=1} ^N\sum_{n_2=1} ^N 
 | {\mathbb E} 
\langle \tilde \phi_{p_1} ^{n_1} ; \tilde \phi_{p_2} ^{n_2}; \tilde \phi_q ^m ; \tilde \phi_{-q}^n \rangle_g |^2 
\leq {\rm Var} \tilde \chi ^{m,n}(q,g) \leq 
{\mathbb E} |
\langle
\tilde \phi_{q} ^m 
\rangle_g |^2.  
\end{eqnarray}
These results calculated in the large $N$ expansion with the replica method agree with these inequalities. 
\section{Summary}
A new series of inequalities for correlation functions in random field systems has been obtained systematically in the square interpolation
which is a mathematically rigorous method. 
The first inequality is the Schwartz-Soffer inequality which gives the relation between  connected and disconnected two-point functions.
This is well-known as a useful inequality to check critical exponents of 
two-point correlation functions calculated in 
effective theories and numerical studies \cite{SS}.
Other inequalities give new relations among multiple-point correlation functions. 
These relations  enable us to examine several critical exponents calculated in large $N$ expansion with the replica method \cite{SMI}.
All obtained results satisfy these inequalities.

\ack
We would like to thank N. G. Fytas for informing us recent numerical studies on 
the random field Ising model in several dimensions.

\end{document}